\documentclass[a4paper,11pt]{article}
\pdfoutput=1

\usepackage{jcappub}
\usepackage[T1]{fontenc}
\usepackage{longtable}

\newcommand{\fluxunit}{\ensuremath{\mathrm{ph\,cm^{-2}\,s^{-1}}}}
\newcommand{\eflux}{\ensuremath{\mathrm{erg\,cm^{-2}\,s^{-1}}}}
\newcommand{\etacar}{$\eta$~Car}
\newcommand{\degr}{\ensuremath{^{\circ}}}

\makeatletter
\def\LT@makecaption#1#2#3{%
  \LT@mcol\LT@cols c{\hbox to\z@{\hss\parbox[t]\LTcapwidth{%
    {\bf #1{#2.}} #3\endgraf\vskip\baselineskip}\hss}}}
\makeatother

\title{\boldmath An 18-year \textit{Fermi}-LAT stacking limit on GeV $\gamma$-ray emission from particle-accelerating colliding-wind binaries}

\author{Youngwan Son}
\affiliation{Institute of Basic Science, Sungkyunkwan University, Suwon, Republic of Korea}
\emailAdd{youngwan.son@cern.ch}

\abstract{The wind-collision regions of massive colliding-wind binaries (CWBs) accelerate relativistic particles, as shown by their non-thermal radio synchrotron and, in \etacar, hard X-ray emission~\cite{hamaguchi2018}. Whether CWBs emit GeV $\gamma$-rays as a population is unknown: only \etacar\ is unambiguously detected by the \textit{Fermi} Large Area Telescope (LAT). We analyze 17.8\,yr of \textit{Fermi}-LAT data above $1$\,GeV (where the sharp point-spread function controls plane confusion) at the 61 confirmed (list~A) particle-accelerating CWBs (PACWBs) of De Becker \& Raucq~\cite{debecker2013}, using a two-dimensional (photon index $\times$ flux) likelihood scan against 200 control fields matched in latitude, local source density, and diffuse intensity. Removing systems whose lines of sight coincide with bright catalogue $\gamma$-ray sources leaves a clean, mutually isolated sample of 6, whose stack is consistent with the control-field null ($p=0.83$): no evidence for collective GeV emission. Retaining those systems instead yields a spurious $\gtrsim3.5\sigma$ excess from chance catalogue coincidences, a caveat for Galactic-plane stacking. The resulting 95\% limit on the mean per-source flux, $F(>1\,\mathrm{GeV})\lesssim1.1\times10^{-11}$~\fluxunit\ ($\Gamma=2$), implies a $\gamma$-ray production efficiency $\eta=L_\gamma/L_{\rm wind}\lesssim4\times10^{-6}\,(d/\mathrm{kpc})^2$, about two orders of magnitude below \etacar\ at the sample's median distance. A representative single-zone model then bounds the electron acceleration efficiency in CWB wind-collision regions unless their magnetic field is comparable to or above the magnetic--photon equipartition value. Among the high-latitude systems accessible to a clean stack, \etacar\ thus appears to be a singular object rather than the brightest member of an emerging population.}


\begin{document}
\maketitle
\flushbottom

\section{Introduction} \label{sec:intro}

Massive binaries whose supersonic stellar winds collide form a wind-collision region (WCR) bounded by strong shocks that can accelerate particles to relativistic energies~\cite{eichler1993}. The defining observational signature is non-thermal radio synchrotron emission, used to compile the catalogue of particle-accelerating colliding-wind binaries (PACWBs)~\cite{debecker2013,debecker2017}. Diffusive shock acceleration in the WCR is expected to produce relativistic electrons and protons, and hence inverse-Compton (IC) and/or neutral-pion-decay $\gamma$-ray emission at GeV to TeV energies~\cite{eichler1993,pittard2006,reitberger2014,pittard2021}.

Despite clear evidence for particle acceleration, the GeV $\gamma$-ray census of CWBs remains sparse. \etacar\ is the only unambiguously \textit{Fermi}-LAT--detected CWB, a high-significance, persistent, catalogued point source positionally coincident with the binary and confirmed across multiple analyses~\cite{abdo2010,reitberger2015,balbo2017,martidevesa2021}. The first dedicated LAT survey (seven CWBs analyzed individually with two years of data) found no significant emission~\cite{werner2013}. With longer exposure $\gamma^2$~Velorum (WR\,11) was detected at $\sim6\sigma$~\cite{pshirkov2016}, with hints of orbital variability~\cite{martidevesa2020}. In the crowded Vela region, however, it is a less secure detection than \etacar. No further CWB has been individually detected. Whether the \emph{population} of CWBs produces a collective, individually sub-threshold GeV signal remains open.

Stacking, which co-adds the likelihood profiles of many individually undetected sources, is the standard tool for such population searches. Here we apply it across the 61 confirmed (list~A) PACWBs of this catalogue, setting its candidate (list~B) systems aside (section~\ref{sec:data}), using 17.8\,yr of LAT data, with a two-dimensional flux--index likelihood scan~\cite{paliya2019}, a cumulative test-statistic estimator~\cite{song2023}, and a control-field null~\cite{henry2024}. Because most of the catalogue lies in confused Galactic-plane fields, the usable clean stack reduces to 6 systems. From these we report a non-detection and the resulting constraints, and we note a methodological caveat specific to Galactic-plane stacking.

\section{Sample and data} \label{sec:data}

Our sample comprises the 61 confirmed (``list~A'') systems in the online catalogue of particle-accelerating colliding-wind binaries~\cite{debecker2013,debecker2017}: massive binaries (or multiple systems, some with multiplicity not yet investigated) whose non-thermal radio emission marks them as particle accelerators, with wind kinetic powers typically $L_{\rm wind}\sim10^{36}$--$10^{37}\,\mathrm{erg\,s^{-1}}$. We exclude the 14 candidate (``list~B'') systems of less certain status. The sample is strongly concentrated toward the Galactic plane (median $|b|\approx0.8\degr$). The full list is given in table~\ref{tab:sample} (appendix~\ref{app:sample}), ordered with the selected (clean) sample first, then the merged systems, the two individually detected CWBs (\etacar\ and $\gamma^2$~Vel), and the catalogue-masked systems. For each system we list the nearest bright source in the \textit{Fermi}-LAT 16-year catalogue (FL16Y)~\cite{fl16y} and its association.

We analyze the first 17.8\,yr of \textit{Fermi}-LAT data (mission elapsed time $239{,}557{,}417$--$801{,}048{,}747$\,s, 2008 August to 2026 May), reducing every target and every control field in exactly the same way so that the two are directly comparable. For each field we select \texttt{P8R3\_SOURCE}-class events~\cite{atwood2009,bruel2018} (\texttt{evclass}~$=128$, \texttt{evtype}~$=3$) over $1$--$500\,\mathrm{GeV}$ within a $15\degr$ region of interest. We work in this hard band because the point-spread function narrows with energy: its $68\%$ containment is $\lesssim0.8\degr$ here, sharp enough to separate a binary from its Galactic-plane neighbours, whereas the broader response below $1$\,GeV leaves the two blended (section~\ref{sec:discussion}). A zenith-angle cut of $z_{\max}=90\degr$ rejects $\gamma$-rays from the bright Earth limb, and the standard good-time-interval selection (\texttt{DATA\_QUAL>0 \&\& LAT\_CONFIG==1}) keeps only data taken in the nominal science configuration. The events are binned into $0.08\degr$ pixels with eight logarithmic energy bins per decade, and the exposure and response are computed with the \texttt{P8R3\_SOURCE\_V3} instrument response functions, applying the energy-dispersion correction to every component except the isotropic template. The background model comprises the FL16Y catalogue source list and two interstellar templates: the Galactic diffuse emission \texttt{gll\_iem\_uw1216\_v13} (released with FL16Y as a smooth spatial and spectral rescaling of the standard \texttt{gll\_iem\_v07} following 4FGL-DR4~\cite{fl16y,ballet2023}) and the isotropic template \texttt{iso\_P8R3\_SOURCE\_V3}. All likelihood fits use \texttt{fermipy} v1.4.0~\cite{wood2017}, built on the \textit{Fermi} Science Tools (\texttt{fermitools}) v2.4.0.

The control sample consists of 200 control fields (CFs) at random sky positions whose $|b|$ distribution matches the full target sample, following Henry et al.~\cite{henry2024}. Positions within $1.0\degr$ of any FL16Y source with $\mathrm{Signif\_Avg}\geq5$, and within self- and sample-exclusion radii, are avoided. Each field is processed identically to a target and defines the null distribution of the stacking statistic. Applying this $1.0\degr$ mask identically to the targets and the control fields is central to the result (section~\ref{sec:discussion}).

\begin{figure}[tbp]
\centering
\includegraphics[width=.8\textwidth]{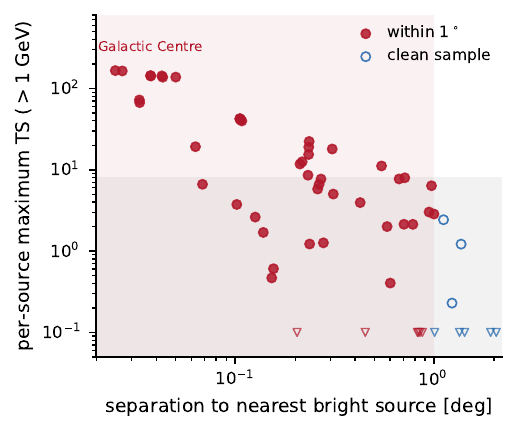}
\caption{Per-source maximum TS ($>1$\,GeV) versus separation to the nearest bright FL16Y source ($\mathrm{Signif\_Avg}\geq5$). Filled red points lie within the $1\degr$ catalogue mask (shaded), open blue points are the clean sample beyond it, and downward triangles mark non-detections ($\mathrm{TS}\le0$) at the floor. The highest-TS coincidences, the Galactic-Center members, are labeled. The two individually detected CWBs (\etacar, $\gamma^2$~Vel), whose nearest bright source is their catalogue counterpart, are omitted. The gray band is the control-field background to its $\approx95^{\rm th}$ percentile: the clean systems fall within it, whereas the catalogue-coincident ones track the unrelated bright source.\label{fig:contam}}
\end{figure}

From these 61 systems we define the analysis sample. The two CWBs that FL16Y catalogues as $\gamma$-ray binaries in their own right (\etacar\ and $\gamma^2$~Vel) are individually detected and set aside, since the search targets the collective signal of the \emph{undetected} population. (In a dedicated single-source fit with that coincident FL16Y counterpart removed, \etacar\ reaches a test statistic $\mathrm{TS}=8098$ ($F(>1\,\mathrm{GeV})=1.3\times10^{-8}$~\fluxunit) and $\gamma^2$~Vel $\mathrm{TS}=74$ (model-dependent in the crowded Vela field, and consistent with the $\sim6\sigma$ of ref.~\cite{pshirkov2016}). The uniform per-source scan (section~\ref{sec:methods}) instead returns only the small residual above this counterpart, which holds most of the flux in our background model, see table~\ref{tab:sample}.) We then apply the FL16Y mask ($1.0\degr$, $\mathrm{Signif\_Avg}\geq5$) to the targets: even above $1$\,GeV the point-spread function (PSF) is broad enough (section~\ref{sec:discussion}) that a PACWB within $\sim1\degr$ of a bright catalogue source is blended with it, and in the plane that source is almost always unrelated: a star-forming region, the Galactic-Center source, or an unassociated detection (table~\ref{tab:sample}, figure~\ref{fig:contam}), so the per-source test would measure the neighbor, not the binary. Finally we require the targets to be mutually separated by at least the CF self-exclusion radius ($1.5\degr$), matching the independent spacing of the control fields that calibrate the null. Targets closer than this, whose LAT fields overlap and would enter the stack as correlated rather than independent positions, are merged into their highest-TS member. In our sample WR\,78 and WR\,79a collapse into HD\,151804 in the Sco~OB1 region. Keeping the highest-TS member rather than averaging biases the stack slightly upward, so the non-detection is conservative. This leaves a clean, mutually isolated sample of $N=6$, a higher-latitude, less crowded subset. The bulk of the catalogue is unsuitable for a clean stack because it lies in confused Galactic-plane fields. \etacar\ is an extreme case: besides its own catalogue counterpart, five unrelated FL16Y sources ($\mathrm{Signif\_Avg}\geq5$) lie within $1\degr$.

\section{Methods} \label{sec:methods}

Individual CWBs are expected to lie well below the LAT detection threshold, so no single system can establish or rule out GeV emission on its own. Stacking addresses this by testing whether the targets share a common, individually sub-threshold signal. Rather than averaging measured fluxes, we combine their full likelihood functions: because the log-likelihood is additive for independent datasets, the joint test statistic of $N$ sources for a common spectral model is the sum of the individual ones,
\begin{equation}
\mathrm{TS}_{\rm stack}(\Gamma,F)=\sum_{i=1}^{N}\mathrm{TS}_i(\Gamma,F)\,,
\end{equation}
with $\mathrm{TS}_i$ the per-source likelihood-ratio statistic. A signal shared by the sample adds coherently, so its expected stacked TS grows as $N$ and the detection significance ($\propto\sqrt{\mathrm{TS}}$) improves by roughly $\sqrt{N}$ relative to a single object, while uncorrelated background fluctuations tend to cancel. This is the standard approach for probing populations of individually undetected $\gamma$-ray sources (see, e.g., refs.~\cite{paliya2019,song2023,henry2024}).

For each target and control field we run a binned likelihood analysis. Starting from the background model of section~\ref{sec:data} (all FL16Y sources within $15\degr$), we optimize the region of interest (ROI), free the diffuse normalizations together with the parameters of significant ($\mathrm{TS}>25$) sources within $5\degr$, and perform a maximum-likelihood fit. Residual point sources are then located iteratively with a TS-map finder down to $\sqrt{\mathrm{TS}}=4$ and added to the model. A residual source coincident with the nominal target position (within its $95\%$ localization radius and $0.2\degr$) is identified as the target itself rather than kept as a separate background component, so the target flux is never split between a background residual and the test source. For the clean sample this never occurs, since every clean system sits well below the finder's $\mathrm{TS}=16$ threshold. We then place a test source (\texttt{PowerLaw2}) at the nominal position and, rather than fix a single spectrum, we scan a grid of photon index $\Gamma\in[1.0,3.5]$ ($\Delta\Gamma=0.1$) and integrated photon flux $\log_{10}[F/\fluxunit]\in[-13,-6]$ (70 bins, its upper bound well above the brightest catalogue-coincident sources, so no source is truncated at the grid edge). At each node the test source is held at that $(\Gamma,F)$ while the Galactic and isotropic diffuse normalizations are re-fitted and all other sources are fixed at their ROI-fit values. Freeing the diffuse at every node guards against the sensitivity of stacked test statistics to the diffuse normalization~\cite{song2023}. We record $\mathrm{TS}(\Gamma,F)=2(\ln\mathcal{L}-\ln\mathcal{L}_{\rm null})$ relative to the source-free model~\cite{paliya2019}. Each source is thus reduced to a two-dimensional TS map on the common $(\Gamma,F)$ grid, and these maps are what we stack.

We combine the maps in two complementary ways. Their sum gives $\mathrm{TS}_{\rm stack}(\Gamma,F)$, whose maximum over the grid is the stacked detection TS for a common spectrum. Because the sample is inhomogeneous in distance, exposure, and contamination, we also use the cumulative-TS statistic of Song et al.~\cite{song2023}, the sum of the per-source maximum TS. Its significance follows from the control-field null below, so we do not require the random-order resampling that Song et al.~\cite{song2023} use to estimate the stack variance. Because the total is also independent of the summation order, we instead form the running sum in order of descending per-source TS. This leaves the significance unchanged and makes the accumulation diagnostic, indicating whether the total is broad-based or carried by a few outliers, as we quantify in section~\ref{sec:gev}.

The null distribution of these statistics is not analytic, since along the Galactic plane it is shaped by diffuse mismodeling and source confusion rather than by photon noise alone. We therefore calibrate it directly from the control fields~\cite{henry2024}. Because the per-source background TS in the plane is set by the local source confusion and diffuse intensity more directly than by $|b|$ alone, our fiducial null matches each clean target to its $K=20$ nearest control fields in a standardized three-covariate space: Galactic latitude $|b|$, the local FL16Y source density (the summed $\mathrm{Signif\_Avg}$ of catalogue sources with $\mathrm{Signif\_Avg}\geq5$ within $2\degr$, the radius enclosing the $95\%$ PSF wing just outside the $1\degr$ mask), and the $>1$\,GeV Galactic-diffuse intensity. We repeat the draw $2\times10^4$ times to build the null distribution and its $\pm2\sigma$ band. Since the draws resample the same 200 fields, the effective size of the null is set by those 200 positions, and the significance is the fraction of realizations whose total equals or exceeds the observed value. Because each control field is reduced through the identical $(\Gamma,F)$ grid maximization applied to the targets, the trials incurred by maximizing over the grid are absorbed into this empirical null. A null matched on $|b|$ alone gives a very similar non-detection, which we report throughout for comparison.

\section{Results} \label{sec:results}

\subsection{The clean-sample stack} \label{sec:gev}

\begin{table}[tbp]
\centering
\begin{tabular}{lc}
\hline\hline
Quantity & Value \\
\hline
Cumulative TS                  & 3.91 \\
CF null median (mean, $\pm2\sigma$) & 10.0 (10.9, $1.0$--$27$) \\
$p$-value (significance)       & 0.83 ($-0.9\sigma$) \\
\quad\textit{($|b|$-only null)} & \textit{0.61 ($-0.3\sigma$)} \\
\hline
\multicolumn{2}{c}{95\% upper limit, clean stack ($\Gamma=2$)} \\
\hline
$F(>1\,\mathrm{GeV})$ [\fluxunit]     & $<1.1\times10^{-11}$ \\
$S(>1\,\mathrm{GeV})$ [\eflux]        & $<1.1\times10^{-13}$ \\
$L_\gamma$ [erg\,s$^{-1}$] $(d/\mathrm{kpc})^{-2}$ & $<1.3\times10^{31}$ \\
$\eta=L_\gamma/L_{\rm wind}$ $(d/\mathrm{kpc})^{-2}$ & $<4\times10^{-6}$ \\
\hline\hline
\end{tabular}
\caption{PACWB population stacking results ($>1$\,GeV, clean sample $N=6$). The significance uses the fiducial environment-matched control-field null (section~\ref{sec:methods}), with the $|b|$-only null listed for comparison. The $95\%$ upper limits assume $\Gamma=2$.\label{tab:results}}
\end{table}

\begin{figure}[tbp]
\centering
\includegraphics[width=.8\textwidth]{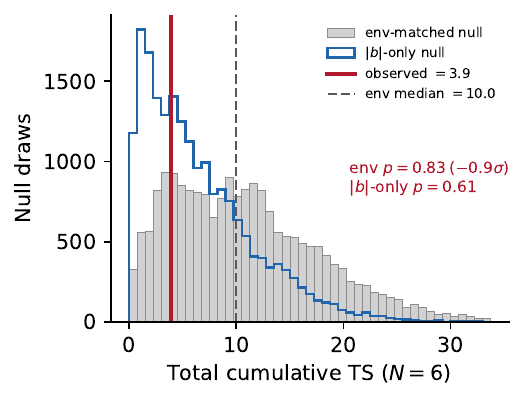}
\caption{Control-field null distribution of the total cumulative TS for the $>1$\,GeV clean stack ($N=6$, $2\times10^4$ realizations). The fiducial environment-matched null (filled) draws each target's control fields from its $K=20$ nearest in $|b|$, local FL16Y source density, and $>1$\,GeV diffuse intensity. The $|b|$-only null is overplotted (blue). The observed value (red, $3.9$) lies below the mode of the right-skewed env-matched null (median $10.0$), giving $p=0.83$ ($-0.9\sigma$). The $|b|$-only null gives $p=0.61$.\label{fig:cumts}}
\end{figure}

The clean sample of $N=6$ has a cumulative TS of $3.91$, consistent with the environment-matched control-field null (figure~\ref{fig:cumts}), with $p=0.83$ ($p=0.61$ for a $|b|$-only null). The null distribution is strongly right-skewed, because a small fraction of control fields fall on chance sub-threshold sources. The observed total therefore lies below the mode and well below the null median of $10.0$ (mean $10.9$, $\pm2\sigma$ band $1.0$--$27$), though still within the bulk of the distribution. We quote significance from the full null distribution rather than from its mean. We find no evidence for collective GeV emission from PACWBs. The cumulative TS is dominated by two targets, 15~Mon and WR\,8 ($\mathrm{TS}\approx2.4$ and $1.2$, together $93\%$ of the total). Both lie below the $90$th percentile of the control-field maximum-TS distribution ($\mathrm{TS}=4.6$), so even the nominally brightest clean members are consistent with background. As an alternative statistic, the summed stacked TS map peaks at $\approx0$ over the $(\Gamma,F)$ grid ($p=0.82$ against the same null), confirming the non-detection. The result is stable against both analysis radii. Tightening the mask to a sub-PSF $0.5\degr$ raises the significance to $p=0.13$ ($1.1\sigma$, section~\ref{sec:robust}), but it also loosens the flux limit to $2.1\times10^{-11}$~\fluxunit\ by re-admitting catalogue-blended systems, so the higher significance is contamination, not signal. Widening the mask to $1.1\degr$ instead (which removes WR\,48, whose nearest bright source sits at $1.004\degr$\footnote{The mask is a strict $\Delta\theta<1\degr$ cut: $\delta$~Ori~A ($0.997\degr$) is masked while the clean WR\,48 ($1.004\degr$) is not, though both round to $1.00\degr$ in table~\ref{tab:sample}.}) leaves the non-detection unchanged ($N=5$, $p=0.72$), so the result is not a mask-boundary artifact. It is also insensitive to the Galactic diffuse background. Restricting the stack to $|b|>2\degr$ leaves the cumulative TS nearly unchanged ($3.7$ for $N=5$, against $3.9$ for all six). The only lower-latitude clean target (HD\,151804, $|b|=1.9\degr$) contributes $\mathrm{TS}=0.2$, so the non-detection is not a diffuse artifact. No clean system is individually significant: the highest reaches only $\mathrm{TS}\approx2.4$ and the archetype WR\,140 is undetected ($\mathrm{TS}\approx0$), all well below the control-field background ($\approx95^{\rm th}$ percentile $\mathrm{TS}=8.2$, maximum $15.1$).

\subsection{Upper limits}

\begin{table}[tbp]
\centering
\small
\begin{tabular}{lrrrrr}
\hline\hline
Name & $\mathrm{TS_{src}}$ & $F_{95}$ & $d$ & $L_\gamma$ & $\eta$ \\
\hline
$\sigma$ Ori AB & 0.0 & 3.3 & 0.44 & 0.8 & $2.9\times10^{-4}$ \\
15 Mon & 2.4 & 8.0 & 0.95 & 8.6 & $9.5\times10^{-4}$ \\
HD 151804 & 0.2 & 11.8 & 1.68 & 39.6 & $2.5\times10^{-4}$ \\
WR 140 & 0.0 & 2.0 & 1.81 & 7.6 & $1.2\times10^{-6}$ \\
WR 48 & 0.0 & 1.9 & 2.40 & 13.3 & $2.3\times10^{-6}$ \\
WR 8 & 1.2 & 9.0 & 3.47 & 129.1 & $3.6\times10^{-5}$ \\
\hline\hline
\end{tabular}
\caption{Individual $95\%$ upper limits for the six clean (list~A) PACWBs ($>1$\,GeV, $\Gamma=2$). $F_{95}$ is in $10^{-11}$\,\fluxunit, $d$ in kpc, and $L_\gamma$ in $10^{31}$\,erg\,s$^{-1}$. $\mathrm{TS_{src}}$ is the per-source maximum over the $(\Gamma,F)$ grid and $F_{95}$ comes from each source's profile likelihood. $d$ is the geometric mean of the catalogue distance range ($L_\gamma\propto d^2$), and $\eta\equiv L_\gamma/L_{\rm wind}$ uses the wind kinetic power from ref.~\cite{debecker2013}, table~A.5.\label{tab:persource}}
\end{table}

We derive a 95\% upper limit on the average per-source flux from the clean $>1$\,GeV stacked profile. At a reference index $\Gamma=2.0$ the stacked TS is consistent with zero (its formal maximum lies at the lowest-flux grid edge), giving
\begin{equation}
F(>1\,\mathrm{GeV}) < 1.1\times10^{-11}~\fluxunit\,,
\end{equation}
or an energy flux $S(>1\,\mathrm{GeV})<1.1\times10^{-13}$~\eflux. The photon-flux limit scales with the assumed index: for $\Gamma=1.5$, $2.0$, $2.5$, and $3.0$ it is $3.4$, $11$, $19$, and $22\times10^{-12}$~\fluxunit, respectively. Results are summarized in table~\ref{tab:results}. The asymptotic $\Delta\mathrm{TS}=2.71$ threshold is conservative here: the env-matched null distribution of the stacked $\Gamma=2$ statistic is narrower than the asymptotic form ($95$th percentile $2.1$), so calibrating to it would tighten the limit by $\approx20\%$. We adopt the conservative asymptotic value as fiducial. The individual $95\%$ limits, from each clean source's own profile, span $F(>1\,\mathrm{GeV})\approx(2$--$12)\times10^{-11}$~\fluxunit, all looser than the stacked average, so the stack sets the tightest per-source constraint (table~\ref{tab:persource}).

The same stacked profile fixes the search depth. Because the stacked TS scales as the square of the flux, a $5\sigma$ detection ($\mathrm{TS}=25$) corresponds to an average flux $\sqrt{25/2.71}\approx3$ times the $95\%$ limit, $F(>1\,\mathrm{GeV})\approx3.3\times10^{-11}$~\fluxunit\ ($\eta\approx1.2\times10^{-5}\,(d/\mathrm{kpc})^2$). A population radiating at \etacar's efficiency ($\eta\sim10^{-3}$) would therefore give a stacked TS far above the null, so the non-detection is a genuine constraint rather than a lack of sensitivity. The limit is averaged over the full 17.8\,yr. The detected CWBs (\etacar\ and $\gamma^2$~Vel) are orbitally modulated, their GeV emission enhanced near periastron~\cite{balbo2017,martidevesa2021}. For a source radiating over a representative $\sim0.2$--$0.3$ of its orbit, the phase-averaged stack would dilute the signal and loosen the per-source limit by $\approx3$--$5\times$, a regime that only phase-resolved modeling (section~\ref{sec:discussion}) could address.

\subsection{Robustness to the analysis radii} \label{sec:robust}

\begin{table}[tbp]
\centering
\begin{tabular}{lcccc}
\hline\hline
Variant & $N_{\rm clean}$ & $p_{\rm env}$ & $p_{|b|}$ & $F_{95}(>1\,{\rm GeV})$ \\
 & & & & (\fluxunit) \\
\hline
\multicolumn{5}{l}{\textit{Merge radius (mask fixed at $1.0\degr$)}}\\
\quad $0.1$--$0.5\degr$ & 8 & 0.92 & 0.80 & $5.9\times10^{-12}$\\
\quad $1.0\degr$ & 6 & 0.83 & 0.61 & $1.1\times10^{-11}$\\
\quad \textbf{$1.5\degr$} & \textbf{6} & \textbf{0.83} & \textbf{0.61} & $\mathbf{1.1\times10^{-11}}$\\
\quad $3.0\degr$ & 6 & 0.83 & 0.61 & $1.1\times10^{-11}$\\
\multicolumn{5}{l}{\textit{Catalogue-mask radius (merge fixed at $1.5\degr$)}}\\
\quad $0.5\degr$ & 18 & 0.13 & 0.02 & $2.1\times10^{-11}$\\
\quad \textbf{$1.0\degr$} & \textbf{6} & \textbf{0.83} & \textbf{0.61} & $\mathbf{1.1\times10^{-11}}$\\
\quad $1.1\degr$ & 5 & 0.72 & 0.55 & $1.8\times10^{-11}$\\
\hline\hline
\end{tabular}
\caption{Robustness of the population non-detection to the two analysis radii, against the fiducial environment-matched null ($p_{\rm env}$) and the $|b|$-only null ($p_{|b|}$). The fiducial analysis (boldface) fixes the FL16Y catalogue-mask radius to $1.0\degr$ and the duplicate-merge radius to $1.5\degr$.\label{tab:robust}}
\end{table}

Table~\ref{tab:robust} shows the population result as the two analysis radii are varied around their fiducial values. The non-detection is insensitive to the duplicate-merge radius and to widening the catalogue mask. Only a sub-PSF mask ($<1\degr$), which re-admits catalogue-blended systems, raises the significance, itself the signature of contamination rather than a population signal.

\section{Discussion} \label{sec:discussion}

\subsection{The efficiency limit and $\eta$~Car}
The per-source flux limit is distance-independent. The implied luminosity and efficiency scale with the assumed distance as $L_\gamma(>1\,\mathrm{GeV})\lesssim1.3\times10^{31}\,(d/\mathrm{kpc})^2\, \mathrm{erg\,s^{-1}}$ and, for a representative wind kinetic power $L_{\rm wind}\sim3.3\times10^{36}\,\mathrm{erg\,s^{-1}}$ ($\dot{M}v_\infty^2/2$ from the catalogue's mass-loss rates and terminal wind speeds), $\eta\equiv L_\gamma/L_{\rm wind}\lesssim4\times10^{-6}\,(d/\mathrm{kpc})^2$. Hence the efficiency bound is per-distance: across the six clean targets, all with catalogue distances ($0.4$--$3.5$\,kpc), the $d^2$ scaling at a representative wind power spans $\eta\lesssim8\times10^{-7}$ (the nearest, $\sigma$~Ori) to $\lesssim5\times10^{-5}$ (the most distant, WR\,8), and is $\lesssim1.1\times10^{-5}$ at the sample median distance of $\sim1.7$\,kpc. The clean systems' catalogue wind powers in fact span $\sim2.7\times10^{34}$--$6.1\times10^{37}\,\mathrm{erg\,s^{-1}}$~\cite{debecker2013}, over three orders of magnitude. The per-source efficiency limit is therefore set mainly by wind power rather than distance, and runs opposite to the distance trend: the strong-winded WR systems give the tightest bounds and the weak-winded O-type systems (such as $\sigma$~Ori) the loosest, source by source in table~\ref{tab:persource}.

\etacar, with $L_\gamma\sim10^{34}$--$10^{35}\,\mathrm{erg\,s^{-1}}$~\cite{abdo2010,martidevesa2021} and a far stronger wind ($L_{\rm wind}\sim10^{38}\,\mathrm{erg\,s^{-1}}$), reaches $\eta\sim10^{-3}$, so the PACWB population is between one and three orders of magnitude less efficient (about two at the median distance) and possibly far lower, since these are upper limits. Given the order-of-magnitude uncertainties in \etacar's $L_\gamma$ and wind power and in the catalogue distances, this efficiency comparison is itself reliable only to about a decade. Its luminosity therefore appears to reflect its own extreme wind power, eccentricity, and periastron density rather than a property shared across the class. No other CWB in the sample approaches this level: with catalogue coincidences removed, the brightest individual targets reach only $\mathrm{TS}\approx2.4$, at the background level, so \etacar\ appears to be a singular object rather than the brightest member of an emerging population. This conclusion holds only for the high-latitude clean subset. Several of the most promising candidate emitters, the closest, highest-wind-power systems in dense star-forming regions (e.g.\ Apep, HD\,93129A, the Cyg~OB2 members, WR\,146), fall in the masked majority and are not probed by the clean stack. A confined population of plane-confused emitters therefore cannot be excluded.

Earlier LAT studies of CWBs analyzed only the seven most promising systems individually and did not stack~\cite{werner2013,pshirkov2016}. Their per-source test statistics in the plane were limited by the soft-band diffuse confusion that our $>1$\,GeV selection and control-field null are designed to control. Their deepest single-source limit, $F(0.1$--$100\,\mathrm{GeV})<1.1\times10^{-9}$~\fluxunit\ for WR\,140~\cite{pshirkov2016}, corresponds to $\approx1\times10^{-10}$~\fluxunit\ above $1$\,GeV for a $\Gamma=2$ spectrum. Our stacked per-source limit lies about an order of magnitude below it, reflecting the deeper exposure, the cleaner band, and the statistical gain of stacking.

\subsection{Constraints on particle acceleration}

\begin{figure}[tbp]
\centering
\includegraphics[width=.8\textwidth]{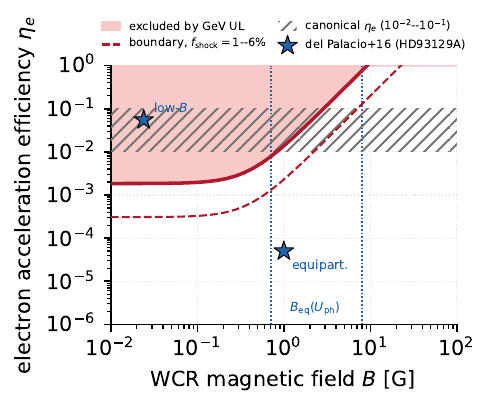}
\caption{Electron-acceleration efficiency $\eta_e$ versus WCR magnetic field $B$ for a representative CWB at $d=1.7$\,kpc (the sample median, $\eta_e^{\rm max}$ scaling as $d^2$). The shaded region is excluded by the population GeV limit (\texttt{naima} inverse Compton with the Klein--Nishina cross section, integrated over $>1$\,GeV). The solid and dashed red curves are the boundary for $f_{\rm shock}=1\%$ and $6\%$, the wind-luminosity fraction through the shock to which $\eta_e$ is normalized~\cite{delpalacio2016}. At the inferred WCR fields ($B\lesssim1$\,G) the limit caps $\eta_e$ at $\sim10^{-3}$. The gray hatched band marks canonical efficiencies ($10^{-2}$--$10^{-1}$), which require $B\gtrsim3$--$7$\,G. Dotted verticals span the magnetic--photon equipartition field $B_{\rm eq}=\sqrt{8\pi U_{\rm ph}}$ (where IC and synchrotron balance) over $U_{\rm ph}=0.02$--$2.5\,\mathrm{erg\, cm^{-3}}$. Stars mark the weak-field and magnetic--thermal equipartition radio solutions of HD\,93129A~\cite{delpalacio2016}, placed at their electron share $\eta_e=f_{\rm NT}/2$ of the quoted non-thermal (electron$+$proton) energy fraction. The weak-field solution lies inside the excluded region, the equipartition solution below it.\label{fig:constraint}}
\end{figure}

The robust result is the population flux limit of section~\ref{sec:results}. Converting it into a statement about particle acceleration requires a source model and is necessarily illustrative. We model a representative CWB with \texttt{naima}~\cite{zabalza2015}: a power-law electron population of index $\alpha=2.2$, close to the test-particle strong-shock value, with an exponential cutoff at $\sim1$\,TeV set by the balance of shock acceleration against synchrotron and inverse-Compton losses~\cite{pittard2021}, radiating inverse Compton (treated as isotropic) on the stellar radiation field, including Klein--Nishina suppression, and synchrotron on the WCR field $B$. Here $B$ is the field of the shocked wind at the collision region, where the electrons radiate. It derives from the stellar surface field advected by the wind, and the synchrotron emission that defines a PACWB already requires $B$ of order mG, far above the $\sim\mu$G interstellar value. For a fixed spectral shape both luminosities scale with the electron energy content, so the GeV limit caps the electron acceleration efficiency at $\eta_e^{\rm max}(B)=\eta_{\rm UL}\,(P_{\rm IC}+P_{\rm sync})/(f_{\rm shock}\,P_{\rm IC,LAT})$, where $\eta_{\rm UL}$ is the GeV efficiency limit, $P_{\rm IC,LAT}$ is the IC power in the LAT band, $P_{\rm IC}$ and $P_{\rm sync}$ are the bolometric IC and synchrotron powers, and $f_{\rm shock}\sim1$--$6\%$ is the fraction of the wind luminosity carried through the shock, to which the acceleration efficiency is conventionally normalized~\cite{delpalacio2016} (figure~\ref{fig:constraint}). This requires a concrete source, for which we adopt a CWB at $d=1.7$\,kpc (the sample median), giving $\eta_{\rm UL}\approx1.1\times10^{-5}$. For a representative hot-star photon field ($L_*\sim5\times10^5\,L_\odot$ at a WCR--star distance of $\sim$10\,AU) and the fields inferred for CWB wind-collision regions ($B$ from a few mG to $\sim1$\,G~\cite{delpalacio2016,pittard2021}), this caps the efficiency at $\eta_e\lesssim10^{-3}$ at the lowest fields, the $f_{\rm shock}$ range giving the factor-of-several band in figure~\ref{fig:constraint}. A canonical efficiency ($\eta_e\sim0.01$--$0.1$, e.g.\ $\eta_e\approx0.055$ for HD\,93129A and up to $\sim$30\% of the shock power into non-thermal particles in CWB models~\cite{delpalacio2016,pittard2021}) would instead require $B\gtrsim3$--$7$\,G, about twice the magnetic--photon equipartition field and above any inferred CWB value.

The radio data alone do not fix $\eta_e$, and the GeV limit breaks the degeneracy. The non-thermal synchrotron flux fixes only the product of the relativistic-electron content and a power of the magnetic field, so radio data admit a continuum of solutions from a weak field with many relativistic electrons to a strong field with few. For HD\,93129A, del Palacio et al.~\cite{delpalacio2016} bracket this continuum with two models: a \emph{weak-field} solution ($B\approx24$\,mG) that puts a large non-thermal fraction through the shock ($f_{\rm NT}\approx0.11$, i.e.\ $\eta_e\approx0.055$ under the assumed electron--proton equipartition), and an \emph{equipartition} solution ($B\approx1$\,G, at which the magnetic energy density equals the thermal) requiring only $f_{\rm NT}\approx10^{-4}$ ($\eta_e\approx5\times10^{-5}$). Inverse Compton, by contrast, scales with the electron content but not with $B$, so the GeV limit caps $\eta_e$ independently of the field: the weak-field solution lies far inside the region excluded for a representative CWB, and only near-equipartition fields survive. (HD\,93129A is itself in our masked sample, so this is a class-level comparison, not an individual limit.) The requirement of a field comparable to or above equipartition is robust to the two least-constrained inputs. The photon field cancels in the inverse-Compton-only bound and moves only the equipartition knee $B_{\rm eq}=\sqrt{8\pi U_{\rm ph}}$ (where IC and synchrotron balance, distinct from the magnetic--thermal equipartition above): over $U_{\rm ph}=0.02$--$2.5\,\mathrm{erg\,cm^{-3}}$, $B_{\rm eq}$ runs $0.7$--$7.9$\,G and the canonical-$\eta_e$ threshold $\approx1.5$--$17$\,G, the two scaling together so that the threshold stays near $2\times$ equipartition. Varying the electron index over $\alpha=2.0$--$3.2$ and the cutoff over $0.3$--$3$\,TeV moves that threshold over $\approx2$--$9$\,G, about $1$--$4$ times the equipartition field. It falls to the equipartition value only for the hardest spectrum with the highest cutoff and is a few times higher otherwise, so a canonical efficiency requires a field comparable to or above equipartition across the range. The fiducial $\alpha=2.2$ lies at the hard end of the slopes indicated by CWB radio spectra and by the soft, two-component GeV spectrum of \etacar~\cite{martidevesa2021}. The softer values those data favor raise the required field rather than lower it.

This mapping is illustrative, not a measurement of the population's fields. It scales with the adopted distance and wind power as $\eta_e^{\rm max}\propto d^2/L_{\rm wind}$, replaces the extended, anisotropic wind-collision region with a single-zone isotropic emitter, and pins the HD\,93129A benchmarks to their electron share under electron--proton equipartition (a proton-dominated ratio would move them downward). A faithful treatment would be per-source and phase-resolved, beyond a stacking analysis. It also assumes the electron spectrum reaches the $\sim$10--100\,GeV energies that inverse-Compton into the LAT band. Since PACWBs are non-thermal radio sources, relativistic electrons are certainly present at lower energies, so the non-detection constrains the acceleration efficiency and the spectral reach jointly: a low efficiency with a hard spectrum, and an ordinary efficiency with a spectrum that cuts off below those energies~\cite{reitberger2017}, read alike.

A hadronic channel is also expected, but the GeV limit constrains it far more weakly. Relativistic protons radiate through $pp\to\pi^0\to2\gamma$ on the wind material. At a representative wind-collision-region density ($n_{\rm H}\sim10^7\,\mathrm{cm^{-3}}$~\cite{pittard2006}) the $pp$ cooling time ($\sim$yr) far exceeds the advection time out of the region ($\sim$weeks), so only a fraction $\sim10^{-3}$--$10^{-2}$ of the proton energy is radiated. Propagating the $>1$\,GeV limit through a \texttt{naima} pion-decay model~\cite{zabalza2015} (a proton spectrum of index $\alpha=2.2$ with a $1$\,PeV cutoff) gives, on the same shock-power normalization used above, a proton efficiency $\eta_p\lesssim0.11$ at this density (loosening to order unity at $10^6$ and tightening to $\sim1\times10^{-2}$ at $10^8\,\mathrm{cm^{-3}}$), comparable to or above canonical efficiencies~\cite{pittard2021}, and more than an order of magnitude weaker than the leptonic bound. Varying the proton index over $\alpha=2.0$--$2.6$, up to the softer slope of the hadronic high-energy component of \etacar\ (the only CWB with a measured GeV spectrum, $\Gamma\approx2.55$~\cite{martidevesa2021}), changes $\eta_p$ by a factor $\lesssim4$ at fixed density, subdominant to the $n_{\rm H}$ uncertainty that dominates the hadronic bound. The limit dips below canonical proton efficiencies ($\eta_p\lesssim0.1$) only for $n_{\rm H}\gtrsim10^7\,\mathrm{cm^{-3}}$, and constrains the full canonical range only for $n_{\rm H}\gtrsim10^8\,\mathrm{cm^{-3}}$. The non-detection therefore constrains efficient electron acceleration but not efficient proton acceleration: the protons escape the wind-collision region before radiating, so the $\gamma$-rays trace only a small fraction of their energy~\cite{pittard2006,reitberger2014}.

\subsection{The control-field method in the Galactic plane}
We fix the catalogue-mask radius a priori to the $1\degr$ used in generating the control fields rather than tuning it. That the significance depends on this radius at all (section~\ref{sec:results}) is itself diagnostic, since a genuine population signal would be insensitive to it. The energy band is fixed a priori on physical grounds. Toward lower energies the LAT point-spread function broadens to several degrees and the bright, structured Galactic diffuse emission leaves per-source test statistics in the plane limited by systematics rather than by photon noise, so no mask cleanly separates a binary from its neighbors. Above $1$\,GeV the PSF core sharpens ($68\%$ containment $\lesssim0.8\degr$), suppressing source confusion and the diffuse systematic. Its wings nonetheless stay broad ($95\%$/$99\%$ containment $\approx2.4\degr$/$4.4\degr$ near $1$\,GeV), so the $1\degr$ mask encloses only $\approx78\%$ of a source's PSF ($\approx72\%$ at $1$\,GeV, where most photons lie). The mask removes the bright core, and the identically masked control fields cancel the residual wings. Any imperfect cancellation, slightly larger for the denser target fields, would inflate the flux limit rather than create a spurious signal, so it acts conservatively on the non-detection. Because the inverse-Compton emission expected from CWBs peaks in the GeV band, the band is also well matched to the target spectrum, at the cost of roughly an order of magnitude fewer photons.

The control-field null absorbs the contamination common to random plane positions, while the catalogue mask and the merging step remove its asymmetric and self-clustering components. The fiducial environment-matched null is robust: the result is stable against the source-density radius ($p=0.83$, $0.80$, $0.76$ for $2\degr$, $3\degr$, $4\degr$) and the neighbor count ($p=0.82$--$0.83$ for $K=10$--$30$), every leave-one-out subsample of five remains a non-detection ($p=0.71$--$0.91$), and bootstrapping the 200 control fields gives $p=0.81\pm0.10$, set by the finite pool rather than the $2\times10^4$ Monte-Carlo draws (whose error is $\pm0.003$). The clean targets in fact lie in slightly denser, higher-diffuse fields than the average control field (directly countering the concern that the control fields are more contaminated than the targets themselves), so the $|b|$-only null ($p=0.61$) is, if anything, the more conservative of the two. The highest-latitude clean target ($\sigma$~Ori) has no catalogue source within $2\degr$, leaving its density covariate inactive, but it carries negligible TS. Two model dependences remain unprobed. We do not vary the \emph{form} of the Galactic diffuse model, the dominant systematic for plane analyses. Freeing its normalization at every grid node (section~\ref{sec:methods}) controls only the first-order sensitivity, though the stability of the stack above $|b|>2\degr$ (section~\ref{sec:gev}), where the diffuse emission is weakest, argues against a diffuse-driven artifact. Our reliance on the recently released FL16Y is likewise confined to the source list that sets the catalogue mask and the local-density covariate. The clean non-detection arises in source-poor, high-latitude fields where the catalogue has little leverage, so it is unlikely to hinge on FL16Y rather than the public 4FGL-DR4.

Our result also points to a requirement for control-field stacking along the Galactic plane: the catalogue mask must be applied identically to the targets and to the control fields. Plane source samples are spatially correlated with catalogued $\gamma$-ray sources, since both trace star-forming regions, so an asymmetric mask produces a spurious population signal from chance coincidences alone. Our data show this directly. Stacking the undetected population (the 59 list~A systems other than \etacar\ and $\gamma^2$~Vel) without the catalogue mask reaches a cumulative TS of $1604$, exceeding all $2\times10^4$ control-field draws of a null centered at $92\pm19$ (an empirical, draw-limited lower bound on the significance, $\gtrsim3.5\sigma$). The excess is driven by a few PACWBs projected on the Galactic-Center Quintuplet/Arches sources, not by collective CWB emission.

The matched $1.0\degr$ mask is conservative and leaves only $N=6$ clean systems (a higher-latitude, less crowded subset of the catalogue), so the residual statistical power is modest and the limit characterizes these uncontaminated systems rather than the plane-confused majority. A smaller mask would recover sources. Building the source-density and diffuse matching used here into the control-field generation itself, rather than applying it through the nearest-neighbor draw, would then keep the null fair under such a mask, a natural extension for a larger sample.

\section{Summary} \label{sec:summary}

We searched 17.8\,yr of \textit{Fermi}-LAT data above $1$\,GeV for collective GeV emission from the 61 confirmed (list~A) PACWBs of De Becker \& Raucq~\cite{debecker2013} via control-field stacking. The apparently significant naive stack is an artifact of catalogue-source coincidences combined with an asymmetric catalogue mask. The clean stack of 6 systems is consistent with the null, yielding a 95\% per-source limit $F(>1\,\mathrm{GeV})\lesssim1.1\times10^{-11}$~\fluxunit\ ($\Gamma=2$) and an efficiency $\eta\lesssim4\times10^{-6}\,(d/\mathrm{kpc})^2$. This caps the electron acceleration efficiency in the wind-collision regions unless their magnetic field is comparable to or above magnetic--photon equipartition. \etacar\ remains the only unambiguously LAT-detected CWB, though the clean stack probes only a high-latitude subset. The most promising candidates lie in confused regions removed by the mask.

\acknowledgments

This work made use of publicly available \textit{Fermi}-LAT data and analysis tools provided by the \textit{Fermi} Science Support Center, operated by NASA's Goddard Space Flight Center, and of the High Energy Astrophysics Science Archive Research Center (HEASARC). The analysis used \texttt{fermipy}~\cite{wood2017}, the \textit{Fermi} Science Tools (\texttt{fermitools}), \texttt{naima}~\cite{zabalza2015}, \texttt{astropy}, \texttt{numpy}, \texttt{scipy}, and \texttt{matplotlib}.

{\sloppy This research was supported by Basic Science Research Program through the National Research Foundation of Korea (NRF) funded by the Ministry of Education (2018R1A6A1A06024977) and the Ministry of Science and ICT (RS-2023-NR076954, RS-2026-25470489, RS-2026-25504820).\par}

\appendix

\section{The stacking sample} \label{app:sample}

{\footnotesize
\setlength{\tabcolsep}{3pt}
\begin{longtable}{lrrrrlcr}
\caption{The 61 confirmed (list~A) particle-accelerating colliding-wind binaries in the stacking sample, ordered as the selected (clean) sample, the spatially merged duplicates, the two individually detected CWBs ($\eta$~Car and $\gamma^2$~Vel, set aside), and the systems removed by the FL16Y catalogue mask. $\mathrm{TS_{src}}$ is each binary's per-source maximum test statistic ($>1$\,GeV). The last three columns give the nearest FL16Y catalogue source at $\mathrm{Signif\_Avg}\geq5$, its catalogue association (`unassoc.' if none), and its angular separation $\Delta\theta$. Only for $\eta$~Car and $\gamma^2$~Vel is that source the binary itself. Systems with $\Delta\theta<1\degr$ are catalogue-coincident and excluded.\label{tab:sample}}\\
\hline\hline
Name & $l$ & $b$ & $d$ & $\mathrm{TS_{src}}$ & Nearest FL16Y src & Association & $\Delta\theta$ \\
 & (deg) & (deg) & (kpc) & ($>1$\,GeV) & & & (deg) \\
\hline
\endfirsthead
\multicolumn{8}{c}{{Table~\thetable, continued from previous page}}\\
\hline\hline
Name & $l$ & $b$ & $d$ & $\mathrm{TS_{src}}$ & Nearest FL16Y src & Association & $\Delta\theta$ \\
 & (deg) & (deg) & (kpc) & ($>1$\,GeV) & & & (deg) \\
\hline
\endhead
\hline
\multicolumn{8}{r}{\textit{continued on next page}}\\
\endfoot
\hline\hline
\endlastfoot
\multicolumn{8}{c}{\textit{Selected (clean) sample, $N=6$}}\\
\hline
15 Mon & 202.9 & +2.2 & 0.9 & 2.4 & J0643.4+0857 & PMN J0643+0857 & 1.11 \\
WR 8 & 247.1 & -3.8 & 3.5 & 1.2 & J0747.3-3310 & PKS 0745-330 & 1.36 \\
HD 151804 & 343.6 & +1.9 & 1.7 & 0.2 & J1645.0-4124 & unassoc. & 1.23 \\
$\sigma$ Ori AB & 206.8 & -17.3 & 0.4 & 0.0 & J0543.0-0420 & unassoc. & 2.05 \\
WR 140 & 80.9 & +4.2 & 1.8 & -0.0 & J2030.9+4416 & PSR J2030+4415 & 1.92 \\
WR 48 & 304.7 & -2.5 & 2.4 & -0.0 & J1311.5-6614 & unassoc. & 1.00 \\
\hline
\multicolumn{8}{c}{\textit{Merged into a clean target ($\Delta\theta<1.5\degr$)}}\\
\hline
WR 78 & 343.2 & +1.4 & 1.6 & 0.0 & J1645.0-4124 & unassoc. & 1.42 \\
WR 79a & 344.1 & +1.5 & 1.6 & -0.0 & J1700.0-4013 & NVSS J165941-401121 & 1.34 \\
\hline
\multicolumn{8}{c}{\textit{Individually detected, set aside}}\\
\hline
$\eta$ Car & 287.6 & -0.6 & 2.4 & 8098$^{\rm a}$ & J1045.0-5941 & Eta Carinae & 0.00 \\
WR 11 & 262.8 & -7.7 & 0.3 & 74$^{\rm a}$ & J0809.4-4715 & Gamma02 Velorum & 0.07 \\
\hline
\multicolumn{8}{c}{\textit{Removed by the catalogue mask ($\Delta\theta<1\degr$), $N=51$}}\\
\hline
Dong66 & 0.2 & -0.1 & 8.2 & 165.8 & J1746.2-2851 & unassoc. & 0.02 \\
Muno7679 & 0.2 & -0.1 & 8.2 & 164.1 & J1746.2-2851 & unassoc. & 0.03 \\
LHO090 & 0.2 & -0.1 & 8.2 & 143.9 & J1746.2-2851 & unassoc. & 0.04 \\
Hos332 & 0.2 & -0.1 & 8.2 & 142.6 & J1746.2-2851 & unassoc. & 0.04 \\
qF257 & 0.2 & -0.1 & 8.2 & 142.2 & J1746.2-2851 & unassoc. & 0.04 \\
qCG1 & 0.2 & -0.1 & 8.2 & 137.8 & J1746.2-2851 & unassoc. & 0.05 \\
qF344 & 0.2 & -0.1 & 8.2 & 137.5 & J1746.2-2851 & unassoc. & 0.04 \\
CEN1a & 15.1 & -0.7 & 2.1 & 72.2 & J1820.4-1609 & NGC 6618 & 0.03 \\
CEN1b & 15.1 & -0.7 & 2.1 & 66.9 & J1820.4-1609 & NGC 6618 & 0.03 \\
FNG2002 26 & 0.1 & +0.0 & 8.2 & 42.5 & J1746.2-2851 & unassoc. & 0.11 \\
WR 102ba & 0.1 & +0.0 & 8.5 & 42.2 & J1746.2-2851 & unassoc. & 0.11 \\
FNG2002 18 & 0.1 & +0.0 & 8.2 & 41.6 & J1746.2-2851 & unassoc. & 0.11 \\
WR 102ah & 0.1 & +0.0 & 8.2 & 41.1 & J1746.2-2851 & unassoc. & 0.11 \\
FNG2002 19 & 0.1 & +0.0 & 8.2 & 39.9 & J1746.2-2851 & unassoc. & 0.11 \\
WR 77o & 339.6 & -0.4 & 4.0 & 22.3 & J1648.4-4554 & unassoc. & 0.24 \\
HD 168112 & 18.4 & +1.6 & 1.7 & 19.2 & J1818.8-1204 & 1RXS J181839.5-12062 & 0.06 \\
Wd1 W15 & 339.6 & -0.4 & 4.0 & 19.0 & J1648.4-4554 & unassoc. & 0.23 \\
WR 98a & 358.1 & -0.0 & 1.9 & 18.0 & J1741.5-3050 & NVSS J174148-305035 & 0.31 \\
Wd1 W17 & 339.6 & -0.4 & 4.0 & 15.4 & J1648.4-4554 & unassoc. & 0.23 \\
WR 105 & 6.5 & -0.5 & 1.6 & 12.5 & J1801.6-2326 & SNR G006.5-00.4 & 0.22 \\
WR 104 & 6.4 & -0.5 & 1.6 & 11.8 & J1801.6-2326 & SNR G006.5-00.4 & 0.21 \\
HD 150136 & 336.7 & -1.6 & 1.3 & 11.1 & J1640.4-4917 & unassoc. & 0.54 \\
Cyg OB2-12 & 80.1 & +0.8 & 1.7 & 8.6 & J2032.2+4127 & PSR J2032+4127 & 0.23 \\
WR 14 & 267.6 & -1.6 & 2.0 & 8.0 & J0859.1-4730 & RCW 38 & 0.71 \\
WR 146 & 80.6 & +0.4 & 1.2 & 7.7 & J2032.2+4127 & PSR J2032+4127 & 0.67 \\
Cyg OB2-9 & 80.2 & +0.8 & 1.7 & 7.7 & J2032.2+4127 & PSR J2032+4127 & 0.27 \\
WR 98 & 355.2 & -0.9 & 1.9 & 6.6 & J1737.2-3331 & unassoc. & 0.07 \\
ALS 15108 AB & 80.1 & +0.7 & 1.7 & 6.6 & J2032.4+4057 & Cyg X-3 & 0.26 \\
HD 15558 & 134.7 & +0.9 & 2.3 & 6.4 & J0240.5+6113A & LSI +61 303 & 0.97 \\
HD 167971 & 18.3 & +1.7 & 1.7 & 5.8 & J1818.8-1204 & 1RXS J181839.5-12062 & 0.26 \\
CD-47 4551 & 268.0 & -1.4 & 1.3 & 5.0 & J0859.1-4730 & RCW 38 & 0.31 \\
Cyg OB2-335 & 80.5 & +0.7 & 1.7 & 3.9 & J2032.2+4127 & PSR J2032+4127 & 0.42 \\
HD 124314 & 312.7 & -0.4 & 1.0 & 3.7 & J1415.8-6143 & WISEA J141501.59-614224.8 & 0.10 \\
Plaskett's star & 205.9 & -0.3 & 1.6 & 3.0 & J0639.4+0655e & Monoceros & 0.94 \\
$\delta$ Ori A & 203.9 & -17.7 & 0.4 & 2.9 & J0529.3-0102 & PMN J0529-0058 & 1.00 \\
HD 93250 & 287.5 & -0.5 & 2.4 & 2.6 & J1045.0-5941 & Eta Carinae & 0.13 \\
Apep & 330.0 & +0.9 & 2.5 & 2.1 & J1601.7-5224 & SNR G329.7+00.4 & 0.70 \\
WR 147 & 79.8 & -0.3 & 0.7 & 2.1 & J2032.7+4009 & NVSS J203231+401619 & 0.78 \\
WR 90 & 343.2 & -4.8 & 1.6 & 2.0 & J1722.7-4531 & SNR G343.0-06.0 & 0.58 \\
HD 93129A & 287.4 & -0.6 & 2.0 & 1.7 & J1043.0-5937 & NGC 3372 & 0.14 \\
WR 21a & 284.5 & -0.2 & 3.0 & 1.3 & J1026.1-5732 & unassoc. & 0.28 \\
Cyg OB2-8A & 80.2 & +0.8 & 1.7 & 1.2 & J2032.2+4127 & PSR J2032+4127 & 0.24 \\
Cyg OB2-5 & 80.1 & +0.9 & 1.7 & 0.6 & J2032.2+4127 & PSR J2032+4127 & 0.16 \\
WR 39 & 290.6 & -0.9 & 5.7 & 0.5 & J1105.3-6108 & PSR J1105-6107 & 0.15 \\
9 Sgr & 6.0 & -1.2 & 1.6 & 0.4 & J1801.8-2358 & NVSS J180200-235857 & 0.60 \\
WR 137 & 74.3 & +1.1 & 1.7 & 0.1 & J2013.7+3614 & SNR G073.9+00.9 & 0.45 \\
WR 133 & 72.7 & +2.1 & 2.1 & 0.1 & J2010.0+3545 & B2 2008+35 & 0.82 \\
WR 125 & 54.4 & +1.1 & 2.0 & 0.0 & J1930.5+1852 & PWN G54.1+0.3 & 0.87 \\
HD 152623 & 344.6 & +1.6 & 1.6 & 0.0 & J1700.0-4013 & NVSS J165941-401121 & 0.85 \\
WR 89 & 348.7 & -0.8 & 3.3 & 0.0 & J1719.8-3856 & NVSS J171936-385653 & 0.20 \\
WR 112 & 12.1 & -1.2 & 4.2 & -0.0 & J1817.3-1947 & WISE G011.662-01.692 & 0.83 \\
\end{longtable}

\noindent $^{\rm a}$ For the two individually detected CWBs the per-source scan value is a residual above their coincident FL16Y counterpart. $\mathrm{TS_{src}}$ here is from a dedicated single-source fit with that counterpart removed ($\eta$~Car: $F(>1\,\mathrm{GeV})=1.3\times10^{-8}$~\fluxunit, $\Gamma=2.3$, and $\gamma^2$~Vel $3.7\times10^{-10}$~\fluxunit\ at $\Gamma=2.6$, the latter model-dependent in the crowded Vela field and consistent with the $\sim6\sigma$ of ref.~\cite{pshirkov2016}). For all other systems it is the per-source maximum of the uniform 2D likelihood scan.
}

\bibliographystyle{JHEP}
\bibliography{refs}
\end{document}